\documentclass[conference, romanappendices,10pt]{IEEEtran}
\IEEEoverridecommandlockouts

\usepackage{amsmath,amssymb,amsthm,bbm,cite,color,enumitem,graphicx,microtype,setspace,url}
\usepackage{graphicx}
\usepackage{tikz}
\usepackage[USenglish]{babel}
\usepackage[utf8]{inputenc}
\usepackage[T1]{fontenc}
\usepackage[algo2e]{algorithm2e}
\usepackage{algorithm}
\usepackage{comment}
\usepackage{epstopdf}

\renewcommand{\a}{\mathbf{a}}

\newcommand{\h}{\mathbf{h}}

\newcommand{\q}{\mathbf{q}}
\renewcommand{\r}{\mathbf{r}}

\newcommand{\w}{\mathbf{w}}

\newcommand{\y}{\mathbf{y}}
\newcommand{\z}{\mathbf{z}}

\newcommand{\A}{\mathbf{A}}

\newcommand{\C}{\mathbf{C}}

\newcommand{\I}{\mathbf{I}}

\newcommand{\setB}{\mathcal{B}}
\newcommand{\setC}{\mathcal{C}}

\newcommand{\setK}{\mathcal{K}}

\newcommand{\setN}{\mathcal{N}}

\newcommand{\Compl}{\mbox{$\mathbb{C}$}}

\newcommand{\Diag}{\mathrm{Diag}}

\newcommand{\Exp}{\mathbb{E}}
\newcommand{\herm}{\mathrm{H}}
\renewcommand{\Im}{\mathrm{Im}}
\newcommand{\maximize}{\mathrm{maximize}}
\newcommand{\minimize}{\mathrm{minimize}}

\renewcommand{\Re}{\mathrm{Re}}
\newcommand{\sgn}{\mathrm{sgn}}
\newcommand{\st}{\mathrm{s.t.}}

\newcommand{\tran}{\mathrm{T}}

\newtheorem{remark}{Remark}

\newcommand{\tar}{\textnormal{\tiny{target}}}

\newcommand{\mse}{\mathrm{MSE}}
\newcommand{\sinr}{\mathrm{SINDR}}
\newcommand{\ue}{\textnormal{\tiny{UE}}}

\usepackage{titlesec} 
\let\subparagraph\relax
\titlespacing{\section}{0pt}{6pt plus 2pt minus 2pt}{5pt plus 1pt minus 2pt}
\titlespacing{\subsection}{0pt}{5pt plus 2pt minus 2pt}{5pt plus 1pt minus 2pt}

\usepackage[labelsep=period,font=small]{caption}
\usepackage{tikz}
\usetikzlibrary{arrows, arrows.meta, backgrounds, calc, intersections, decorations.markings, decorations.pathreplacing, positioning, shapes}

\usepackage{pgfplots}
\usepgfplotslibrary{groupplots}
\pgfplotsset{compat=1.17}

\title{Uplink Power Control for Distributed Massive MIMO with 1-Bit ADCs \vspace{-3mm}}
\author{
\IEEEauthorblockN{Bikshapathi Gouda, Italo Atzeni, and Antti Tölli}
\IEEEauthorblockA{Centre for Wireless Communications, University of Oulu, Finland \\
Emails: \{bikshapathi.gouda, italo.atzeni, antti.tolli\}@oulu.fi \vspace{-2mm}
\thanks{This work is supported by the Academy of Finland (336449 Profi6, 346208 6G~Flagship, 348396 HIGH-6G, and 357504 EETCAMD) and by the European Commission (101095759 Hexa-X-II).}}\vspace{-7mm}}
\IEEEoverridecommandlockouts\IEEEpubid{\makebox[\columnwidth]{ 979-8-3503-1090-0/23/\$31.00~\copyright~2023 IEEE \hfill} \hspace{\columnsep}\makebox[\columnwidth]{ }}
\begin{document}

\maketitle

\begin{abstract}
We consider the problem of uplink power control for distributed massive multiple-input multiple-output systems where the base stations (BSs) are equipped with 1-bit analog-to-digital converters (ADCs). The scenario with a single user equipment (UE) is first considered to provide insights into the signal-to-noise-and-distortion ratio (SNDR). With a single BS, the SNDR is a unimodal function of the UE transmit power. With multiple BSs, the SNDR at the output of the joint combiner can be made unimodal by adding properly tuned dithering at each BS. As a result, the UE can be effectively served by multiple BSs with 1-bit ADCs. Considering the signal-to-interference-plus-noise-and-distortion ratio (SINDR) in the multi-UE scenario, we aim at optimizing the UE transmit powers and the dithering at each BS based on the min-power and max-min-SINDR criteria. To this end, we propose three algorithms with different convergence and complexity properties. Numerical results show that, if the desired SINDR can only be achieved via joint combining across multiple BSs with properly tuned dithering, the optimal UE transmit power is imposed by the distance to the farthest serving BS (unlike in the unquantized case). In this context, dithering plays a crucial role in enhancing the SINDR, especially for UEs with significant path loss disparity among the serving BSs.  

\end{abstract}

\section{Introduction}
Fully digital massive multiple-input multiple-output (MIMO) is widely recognized for its ability to realize flexible beamforming and large-scale spatial multiplexing~\cite{Mar10}. As we move towards higher carrier frequencies in the next-generation wireless systems, the number of antennas at the base station (BS) is bound to increase significantly~\cite{Raj20}. Fully digital massive MIMO arrays with high-resolution analog-to-digital converters (ADCs), however, are exceedingly complex and power hungry. Low-resolution and 1-bit ADCs can be used to considerably decrease the power consumption without compromising the performance~\cite{Jac17,Mol17,Li17,Atz22}. Moreover, adopting low-resolution and 1-bit ADCs in a distributed/cell-free massive MIMO setting allows to reduce the backhaul signaling overhead (among the BSs or between the BSs and the central processing unit) associated with the network-wide beamforming design~\cite{Ngo17,Atz21}.

The works~\cite{Mol17,Li17} investigated the uplink channel estimation and achievable signal-to-interference-plus-noise-and-distortion ratio (SINDR) using 1-bit ADCs in single-BS systems. In addition, the spectral and energy efficiency of massive MIMO systems with low-resolution ADCs are studied as a function of the UE transmit power in~\cite{Pir18}. 
These studies employ the Gaussian approximation for quantization distortion (QD) to evaluate system performance, which accurately represents the behavior in scenarios with low UE transmit power or a large number of UEs. The performance of 1-bit ADC systems with a low number of UEs is evaluated in~\cite{Jac19}, which is inferior at high UE transmit power due to QD. However, the introduction of properly tuned dithering at the BS enhances performance at high transmit powers~\cite{Jac19}. Regarding cell-free massive MIMO systems, the performance of 1-bit ADCs is evaluated in~\cite{Zha19}, while the energy efficiency of such systems with low-resolution ADCs is analyzed in conjunction with an uplink power control scheme for max-min fairness in~\cite{Zha21}. It is important to note that the works in~\cite{Zha19, Zha21} also adopt the Gaussian approximation for QD, which is more accurate with increasing the number of ADC bits. However, the approximation may not fully capture the true performance characteristics of 1-bit ADC systems.

In this paper, we consider the uplink power control problem in the distributed massive MIMO systems with 1-bit ADCs. We first provide insight into the signal-to-noise-and-distortion ratio (SNDR) of a UE in single and multi-BS scenarios with 1-bit ADCs. Our analysis reveals that the SNDR of the UE is non-monotonic and highly dependent on the UE transmit powers and location. Specifically, in the single-BS case, the SNDR is unimodal, whereas, in the multi-BS case, the SNDR is non-unimodal. However, we show that by tuning the Gaussian dithering (i.e., the noise level), the SNDR can be made unimodal even in the multi-BS scenario. We also optimize the transmit powers of UEs by considering the min-power and max-min-SINDR optimization problems in a multi-UE scenario. To solve these optimization problems, we use gradient, fixed-point, and block coordinate descent (BCD) methods. Our numerical results indicate that in a single BS system, the SINDR reaches a saturation point beyond a certain power of the UEs. Moreover, in a multi-BS scenario, if achieving the desired target SINDR involves reception from multiple BSs using a large number of antennas and dithering, then the UE transmit power heavily depends on the distance to the farthest serving BS. In this context, employing dithering at the BSs becomes crucial in enhancing the SINDR for UEs that exhibit a substantial variation in path loss among the serving BSs.

\section{System Model}

We consider an uplink distributed/cell-free massive MIMO system where a set of BSs $\setB \triangleq \{1, \ldots, B\}$ with $M$ antennas serves a set of single-antenna UEs $\setK \triangleq \{1, \ldots, K\}$. Each BS antenna is connected to two 1-bit ADCs for the in-phase and quadrature components of the received signal. Let $\h_{b,k} \in \Compl^{M \times 1}$ denote the uplink  \addtolength{\topmargin}{1mm} channel between UE~$k \in \setK$ and BS~$b \in \setB$, where $\h_{k} \triangleq [\h_{1,k}^{\tran}, \ldots, \h_{B,k}^{\tran}]^{\tran} \in \Compl^{B M \times 1}$ is the aggregated uplink channel of UE~$k$ across all the BSs. 
We use $d_k \sim \setC \setN (0, 1)$ to denote the data symbol transmitted by UE~$k$. The received signal at BS~$b$ prior to the 1-bit ADCs is given by
\begin{align}
\y_b \triangleq \sum_{k \in \setK} \sqrt{\rho_k} \h_{b,k} d_k + \z_b \in \Compl^{M \times 1},
\end{align}
where $\rho_k$ is the transmit power of UE~$k$ and $\z_b \sim \setC \setN (0, \sigma_{b}^{2} \I_{M})$ is the additive white Gaussian noise (AWGN) vector at BS~$b$. Then, $\y_b$ is quantized as
\begin{align}
\r_b \triangleq Q(\y_b) \in \Compl^{M \times 1},
\end{align}
where $Q(\a) \triangleq \frac{1}{\sqrt{2}}\big(\sgn \big(\Re[\a]\big) + j \, \sgn \big(\Im[\a]\big)\big)$. 

Let us introduce $\y \triangleq [\y_1^{\tran}, \ldots, \y_B^{\tran}]^{\tran} \in \Compl^{BM \times 1}$ and $\z \triangleq [\z_1^{\tran}, \ldots, \z_B^{\tran}]^{\tran} \in \Compl^{BM \times 1}$ as the aggregated received signal and AWGN vector, respectively, across all the BSs. Based on the Bussgang decomposition~\cite{Bus52}, we can write
\begin{align} \label{eq:r_vec}
\r \triangleq Q(\y) = \A \y + \q \in \Compl^{BM \times 1},
\end{align}
where $\q$ is the zero-mean, non-Gaussian quantization distortion vector that is uncorrelated with $\y$ and
\begin{align}
    \A \triangleq \sqrt{\frac{2}{\pi}}\Diag(\C_{\y})^{-\frac{1}{2}} \in \Compl^{BM \times BM},
\end{align}
is the Bussgang gain, with $\C_{\y} \triangleq \Exp[\y\y^{\herm}] = \sum_{k \in \setK} {\rho_k} \h_{k} \h_k^{\herm} + \C_\z \in \Compl^{BM \times BM}$ and $\C_\z \triangleq \Diag (\sigma^2_1, \ldots, \sigma^2_B) \otimes \I_{M}$.

The soft-detected symbol of UE~$k$ is obtained by combing $\r$ in \eqref{eq:r_vec} with the combining vector $\w_k \in \Compl^{BM \times 1}$, i.e.,
\begin{align}
    \hat d_k & = \w_k^{\herm} \r \nonumber \\
             & = \sqrt{\rho_k} \w_{k}^{\herm} \A \h_{k} d_k \! + \! \sum_{\bar k \neq k} \! \sqrt{\rho_{\bar k}} \w_{k}^{\herm} \A \h_{\bar k} d_{\bar k} \! + \! \w_{k}^{\herm} \A \z \! + \! \w_{k}^{\herm} \q.
\end{align}
Then, the SINDR of UE~$k$ is given by
\begin{align}\label{eq:sinr_k}
    \sinr_k \! \triangleq \! \frac{\rho_k \|\w_k^{\herm} \A \h_k \|^2}{\sum_{\bar k \neq k} \rho_{\bar k} \|\w_k^{\herm} \A \h_{\bar k} \|^2 \! + \! \| \w_k^{\herm} \A \C_{\z}^{\frac{1}{2}} \|^2 \! + \! \w_k^{\herm} \C_{\q} \w_k},
\end{align}
where  $\C_{\q} \triangleq \Exp [\q \q^{\herm}] = \C_\r - \A \C_{\y} \A^{\herm} \in \Compl^{BM \times BM}$ and~\cite{Li17}
\begin{align}
    \C_\r  & \triangleq \frac{2}{\pi} \big( \arcsin \big( \Diag(\C_{\y})^{-\frac{1}{2}} \Re[\C_\y] \Diag(\C_{\y})^{-\frac{1}{2}} \big) \nonumber \\ 
    & \phantom{=} \ + j \, \arcsin \big( \Diag(\C_{\y})^{-\frac{1}{2}}\Im[\C_\y] \Diag(\C_{\y})^{-\frac{1}{2}} \big) \big).
\end{align}
Finally, the mean squared error (MSE) of UE~$k$ is given by
\begin{align}
    \mse_k \triangleq \Exp \big[ |\w_k^{\herm} \r - d_k|^2 \big].
\end{align}
The combining vector that minimizes the above MSE corresponds to the minimum mean squared error (MMSE) combiner, i.e.,
\begin{align}\label{eq:mmse_wk}
    \w_k = \sqrt{\rho_k} \C_{\r}^{-1} \A \h_k.
\end{align}
By plugging \eqref{eq:mmse_wk} into \eqref{eq:sinr_k}, the latter can be simplified as
\begin{align}\label{eq:sinr_mse_k}
     \sinr_k = \frac{\rho_k \h_k^{\herm}\A \C_{\r}^{-1}\A \h_k }{1-\rho_k \h_k^{\herm}\A \C_{\r}^{-1}\A \h_k }.
\end{align}
In the next section, we will explore the SNDR for a single UE in both single-BS and two-BS scenarios. It is important to note that in the absence of interference from other UEs, the term SINDR can be used interchangeably with SNDR. 

\section{SNDR in the Single-UE Case} \label{sec:SINDRChar}

\begin{figure*}
\vspace{1mm}
    \begin{minipage}[t]{0.31\textwidth}
        \begin{figure}[H]
            \begin{tikzpicture}

\begin{axis}[
	width=6.15cm,
	height=5.15cm,
	xmin=-20, xmax=60,
	ymin=0, ymax=20,
    xlabel={$\rho_{k}$ [dBm]},
    ylabel={$\mathrm{SNDR}$},
    xlabel near ticks,
	ylabel near ticks,
    x label style={font=\footnotesize},
	y label style={font=\footnotesize},
    ticklabel style={font=\footnotesize},
    legend pos=north east,
    legend cell align=left,
    legend columns=1,
    legend style={font=\scriptsize, inner sep=1pt},
	grid=both,
]

\addplot[line width=1pt, black]
table[x=Var1, y=Var2, col sep=comma] 
{Figures/Data/Fig1.txt};

\begin{scope}[>=latex]
\draw[->] (axis cs:10,5) -- (axis cs:-6.5,10) {};
\draw[->] (axis cs:30,15) -- (axis cs:15,10) {};
\end{scope}
\node[align=center] at (axis cs:10,3) {\scriptsize{AWGN} \\[-3pt] \scriptsize{dominates}};
\node[align=center] at (axis cs:30,17) {\scriptsize{quantization distortion} \\[-3pt] \scriptsize{dominates}};

\end{axis}

\end{tikzpicture}
             \vspace{-2mm}
            \caption{\small{Single-BS case: SNDR vs UE transmit power.}} \label{fig:sinBSSINDR}
        \end{figure}
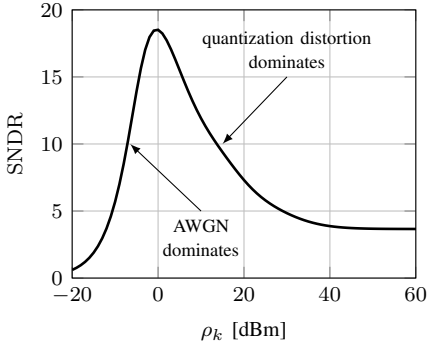
    \end{minipage}
    \hfill
    \begin{minipage}[t]{0.31\textwidth}
        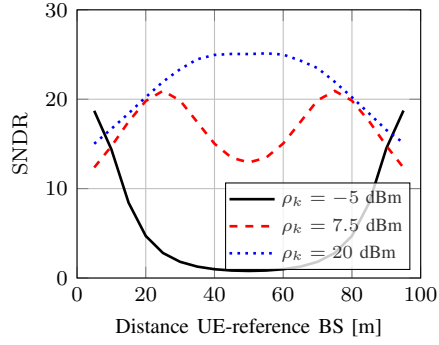
\begin{figure}[H]
            \begin{tikzpicture}

\begin{axis}[
	width=6.15cm,
	height=5.15cm,
	xmin=0, xmax=100,
	ymin=0, ymax=30,
    xlabel={Distance UE-reference BS [m]},
    ylabel={$\mathrm{SNDR}$},
    xlabel near ticks,
	ylabel near ticks,
    x label style={font=\footnotesize},
	y label style={font=\footnotesize},
    ticklabel style={font=\footnotesize},
    legend pos=south east,
    legend cell align=left,
    legend columns=1,
    legend style={font=\scriptsize, inner sep=1pt, fill opacity=0.8, draw opacity=1, text opacity=1},
	grid=both,
]

\addplot[line width=1pt, black]
table[x=dist, y=SNDR_5, col sep=comma] 
{Figures/Data/Fig2.txt};
\addlegendentry{$\rho_{k} = -5$~dBm};

\addplot[line width=1pt, red, dashed]
table[x=dist, y=SNDR_7p5, col sep=comma] 
{Figures/Data/Fig2.txt};
\addlegendentry{$\rho_{k} = 7.5$~dBm};

\addplot[line width=1pt, blue, dotted]
table[x=dist, y=SNDR_20, col sep=comma] 
{Figures/Data/Fig2.txt};
\addlegendentry{$\rho_{k} = 20$~dBm};

\end{axis}

\end{tikzpicture}
             \vspace{-2mm}
            \caption{\small{Two-BS case: SNDR vs distance UE-reference BS.}} \label{fig:2BSSINDR}
        \end{figure}
    \end{minipage}
    \hfill
    \begin{minipage}[t]{0.31\textwidth}
        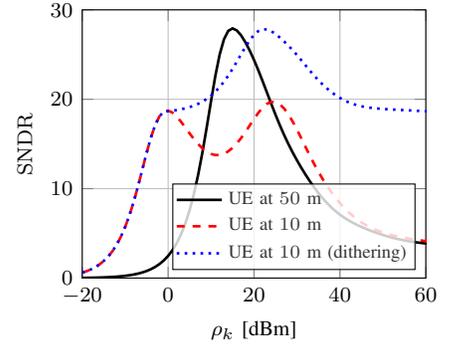
\begin{figure}[H]
            \begin{tikzpicture}

\begin{axis}[
	width=6.15cm,
	height=5.15cm,
	xmin=-20, xmax=60,
	ymin=0, ymax=30,
    xlabel={$\rho_{k}$ [dBm]},
    ylabel={$\mathrm{SNDR}$},
    xlabel near ticks,
	ylabel near ticks,
    x label style={font=\footnotesize},
	y label style={font=\footnotesize},
    ticklabel style={font=\footnotesize},
    legend pos=south east,
    legend cell align=left,
    legend columns=1,
    legend style={font=\scriptsize, inner sep=1pt, fill opacity=0.8, draw opacity=1, text opacity=1},
	grid=both,
]

\addplot[line width=1pt, black]
table[x=pwrdBm, y=UEat50, col sep=comma] 
{Figures/Data/Fig3.txt};
\addlegendentry{UE at $50$~m};

\addplot[line width=1pt, red, dashed]
table[x=pwrdBm, y=UEat10, col sep=comma] 
{Figures/Data/Fig3.txt};
\addlegendentry{UE at $10$~m};

\addplot[line width=1pt, blue, dotted]
table[x=pwrdBm, y=UEat10wTun, col sep=comma] 
{Figures/Data/Fig3.txt};
\addlegendentry{UE at $10$~m (dithering)};

\end{axis}

\end{tikzpicture}
             \vspace{-2mm}
            \caption{\small{Two-BS case: SNDR vs UE transmit power.}} \label{fig:2BSSINDRAWGNtun}
        \end{figure}
    \end{minipage}
    \vspace{-2mm}
\end{figure*}

We consider a single UE and simple network deployments with one or two BSs to characterize the SNDR with respect to the UE transmit power and distance from the BSs. To this end, we provide illustrative numerical results assuming $M=128$ antennas at each BS and employing MMSE combiner. Further, we model the channel between UE $k$ and each BS $b$ as uncorrelated Rayleigh fading, where the channel coefficients are denoted by $\h_{b,k}$. Specifically, the entries of $\h_{b,k}$ are independent and identically distributed (i.i.d.) complex Gaussian random variables with a mean of zero and variance $\delta_{b,k}$. The large-scale fading coefficient $\delta_{b,k}$ is determined by the distance $d_{b,k}$ between BS $b$ and UE $k$, and it is given by $\delta_{b,k} \triangleq -61 - 30\log_{10}(d_{b,k})$ [dB], with the carrier frequency being $28$ GHz and the pathloss exponent being 3. 

\subsection{Single-BS Case: SNDR Versus UE Transmit Power}

Let us first consider a scenario with a single BS equipped with 1-bit ADCs, where the UE is located at a fixed distance of $30$~m from the BS. Fig.~\ref{fig:sinBSSINDR} illustrates the SNDR as a function of the UE transmit power. It can be observed that the SNDR exhibits unimodal behavior, with an optimal value achieved at a specific UE transmit power. At low UE transmit powers, the impact of AWGN dominates, overpowering the intended signal. Conversely, at high UE powers, the received soft symbols suffer from amplitude loss, resulting in QD prevailing over the intended signal. Therefore, the optimal SNDR is achieved by the appropriate amount of AWGN, which induces a beneficial scrambling effect on the 1-bit quantized signals at the BS.

\subsection{Two-BS Case: SNDR Versus Distance UE-Reference BS}\label{subsec:2BSsDis}
We will characterize the SNDR of a UE in a two-BS system. To begin with, the SNDR of a UE is characterized as a function of the distance between the UE and one of the two BSs, which is referred to as the \textit{reference BS} in Fig.~\ref{fig:2BSSINDR}. In this scenario, the two BSs are equipped with 1-bit ADCs and are positioned $100$~m apart. The UE maintains a fixed transmit power and moves along the line connecting the two BSs. At low UE transmit power of $-5$~dBm, the SNDR decreases as the UE moves away from the left BS and increases as it gets closer to the other BS as in the unquantized case, where SNDR is primarily affected by AWGN. At a higher UE transmit power of $7.5$~dBm, SNDR is limited by the QD of the closest BS and AWGN of the farthest BS. However, as the UE moves away from the closest BS, both QD and AWGN impacts reduce, improving SNDR. At the cell edge, i.e., at $50$~m, SNDR is dominated by AWGN at both BSs due to the limited UE transmit power, resulting in lower SNDR. At a very high transmit power of $20$~dBm, SNDR is limited by QD if the UE is close to any of the BSs, and the impact of QD reduces as the UE moves towards the cell edge, resulting in better SNDR. Note that the UE is not power-limited even at the cell edge in this scenario.

\subsection{Two-BS Case: SNDR Versus UE Transmit Power}
Lastly, let us consider the same setup as discussed in Section~\ref{subsec:2BSsDis} to analyze the SNDR with the UE transmit power at a fixed location. Additionally, we also evaluate the impact of dithering at each Base Station (BS). If the UE is equidistant from both BSs, specifically located $50$~m away from the reference BS, the impact of AWGN and QD on the SNDR is the same at both BSs for a fixed UE transmit power level. As a result, the optimal performance at both the BSs is achieved at the same UE transmit power, leading to a unimodal SNDR behavior, as shown in Fig.~\ref{fig:2BSSINDRAWGNtun}, which is similar to the single-BS case. However, the SNDR in the two-BS scenario is higher than in the single-BS case (see Fig~\ref{fig:sinBSSINDR}) due to the combining gain of both the BSs. On the other hand, if the UE is close to one of the BSs (say $10$~m), the optimal performance at the closest BS is achieved with a very low UE transmit power compared to the farthest BS. This results in two peaks in the SNDR behavior, making it non-unimodal. Additionally, adjusting the UE transmit power does not make the SNDR behavior unimodal. Therefore, to enhance the overall SNDR, we can modify the dominance of QD at the closest BS by introducing a dithering signal. In this context, we consider a Gaussian dithering signal, which can be achieved by increasing the power of the AWGN. For simplicity, we consider the parameter $\sigma_b$ as a controllable parameter referred to as the \textit{noise level} parameter. It allows us to adjust the power of the dithering signal. By adjusting the noise level, we can maintain a fixed QD (as mentioned in Remark 1), resulting in an improved SNDR with unimodal behaviour, as shown in Fig.~\ref{fig:2BSSINDRAWGNtun}. In scenarios involving over two BSs and a UE with unlimited power, dithering levels are applied to all BSs except the farthest one. The level of dithering at each BS depends on its pathloss ratio to the farthest BS.

\begin{remark}\rm{
By adjusting the noise level of the 1-bit ADC BS, the QD can be made the same for varying UE transmit powers.}
\end{remark}
To provide a better understanding of Remark 1, we consider a system with a single 1-bit ADC. Let $\sqrt{\rho_d}d$ and $\sigma_z z$ be the input signal and noise to the 1-bit ADC, where $\rho_d$ and $\sigma_z^2$ are the signal power and noise power, respectively. The output of the 1-bit ADC can be expressed as:
\begin{align}\label{eq:r}
r \triangleq \sgn (\sqrt{\rho_d}d+{\sigma_z}z) = \sgn \bigg(\sqrt{\rho_d} \bigg(d+\frac{{\sigma_z}}{\sqrt{\rho_d}}z\bigg)\bigg).
\end{align}
It is clear from \eqref{eq:r} that as long as the ratio $\frac{{\sigma_z}}{\sqrt{\rho_d}}$ is constant, the output of the 1-bit ADC remains the same. Moreover, it can be shown that $\A\y$ in \eqref{eq:r_vec},  does not change if ${\sigma_z}$ and $\sqrt{\rho_d}$ vary with a fixed ratio. Therefore, by adjusting the noise level, i.e., $\sigma_z$, we can fix the $\frac{{\sigma_z}}{\sqrt{\rho_d}}$ ratio for different input signal powers, ensuring that the QD remains the same.

\section{Uplink Power Control}

In this section, we discuss the optimization of the UE transmit powers to achieve a target SINDR and maximize the minimum SINDR. In addition, we can optimize the noise level at the BSs to achieve higher SINDR. Let us begin with the optimization of UE transmit powers using the min-power approach for a fixed target SINDR.

\subsection{Power Control Based on Min-Power}\label{subsec:pcmp}

For a given target SINDR $\gamma_k^{\tar}$ for UE~$k$, the min-power optimization problem can be formulated as follows:
\begin{align}
\begin{array}{cl}
\label{eq:p1} \displaystyle \underset{\{\rho_{k}, \sigma_b\}}{\minimize} & \displaystyle \sum_{k \in \setK} \rho_{k} \\
\st & \sinr_k \ge \gamma_k^{\tar}, \quad \forall k \in \setK \\
    & \sigma_{b} \ge \sigma_{\min}, \quad \forall b \in \setB
\end{array}
\end{align}
where $\sinr_k$ is given in \eqref{eq:sinr_mse_k}, and 
$\sigma^2_{\min}$  is the minimum noise level corresponding to the AWGN power. It is to be noted that the optimization of $\sigma_{b}$ is not required for the single BS case. In the multi-BS case, the noise level at each BS and the UE transmit powers can be tuned in an alternating manner. However, for simplicity, we will first fix the noise levels at the BSs and optimize the UE transmit powers with the following simplified optimization problem:\footnote{\label{note1}For fixed UE transmit power,  a similar optimization problem can be written for noise level optimization in the multi-BS scenario. However, we ignore the noise level optimization for the sake of simplicity and due to space constraints.}
\begin{align} \label{eq:p2}
\hspace{-2mm} \begin{array}{cl}
\displaystyle \underset{\{\rho_{k}\}}{\minimize} & \displaystyle \sum_{k \in \setK} \rho_{k} \\
\st & \displaystyle \rho_k \h_k^{\herm}\A \C_{\r}^{-1} \A  \h_k \ge \frac{\gamma_k^{\tar}}{\gamma_k^{\tar}+1}, \quad \forall k \in \setK.
\end{array}
\end{align}
The constraint in \eqref{eq:p2} on the left-hand side exhibits a unimodal behavior similar to that of $\sinr_k$ in \eqref{eq:sinr_mse_k}. This is because the reformulation of the constraint involves only the addition of a constant value of 1.

The SINDR of a UE with 1-bit is non-monotonic and inherently bounded, unlike the signal-to-interference-plus-noise ratio (SINR) in the unquantized case, even without interference from other UEs. Therefore, existing approaches cannot be used as is to optimize UE transmit powers~\cite{Wie06}. To address this, we propose the gradient, fixed-point, and BCD methods to optimize the powers of \eqref{eq:p2}. 


\begin{figure*}
\addtocounter{equation}{+2}
\begin{align}
\label{eq:gradLag} \frac{\partial}{\partial \rho_k} \mathcal{L}(\{\rho_k, \mu_k\}) & = 1 - \mu_k  \h_k^{\herm}\A \C_{\r}^{-1} \A  \h_k  - \sum_{\bar k \in \setK} \mu_{\bar k} \rho_{\bar k}  \frac{\partial}{\partial \rho_k} \bigg( \h_{\bar k}^{\herm} \A \C_{\r}^{-1} \A \h_{\bar k}\bigg), \\
\label{eq:gradKbar} \frac{\partial}{\partial \rho_k} \h_{\bar k}^{\herm}\A \C_{\r}^{-1} \A  \h_{ \bar k} & =  2 \Re \bigg[\h_{\bar  k}^{\herm} \bigg(\frac{\partial}{\partial \rho_{k}}\A \bigg) \C_{\r}^{-1} \A  \h_{\bar  k} \bigg] -  \h_{\bar k}^{\herm}\A \C_{\r}^{-1} \bigg( \frac{\partial}{\partial \rho_{ k}}  \C_{\r}  \bigg) \C_{\r}^{-1}\A  \h_{ \bar k}
\end{align}
\hrule
\addtocounter{equation}{+1}
\begin{align}\label{eq:gradCrReal}
\frac{\partial}{\partial \rho_k}\Re[\C_{\r}(i,j)] & = \frac{\A(i,i)\A(j,j)}{\sqrt{1-|q_{i,j}|^2}} \bigg( \Re\big[\h_k(i)\h_k(j)\big] - \frac{1}{2}\Re\big[\C_{\y}(i,j)\big] \big( |\h_k(i)|^2 \A^{2}(i,i) +  | \h_k(j)|^2 \A^{2}(j,j) \big)\bigg)
\end{align}
\hrule
\addtocounter{equation}{-6}
\vspace{-3mm}
\end{figure*}

\smallskip

\textit{\textbf{1) Gradient Update.}} Let us begin by writing the Lagrangian of the optimization problem \eqref{eq:p2} as
\begin{equation}
\begin{aligned}
\label{eq:p3Lag}
\mathcal{L}({\rho_k, \mu_k}) \! \triangleq \! \sum_{k \in \setK} \bigg( \! \rho_{k} - \mu_k \bigg( \! \rho_k \h_k^{\herm}\A \C_{\r}^{-1} \A \h_k \! - \frac{\gamma_k^{\tar}}{\gamma_k^{\tar}+1}\bigg)\bigg),
\end{aligned}
\end{equation}
where $\mu_k \ge 0$ is the Lagrangian dual variable associated with the constraint. Similar to Section~\ref{sec:SINDRChar}, it can be observed that $\mathcal{L}({\rho_k, \mu_k})$ is unimodal with respect to $\{\rho_k\}_{ k \in \setK}$ for any $\mu_k \in \mathbb{R}_{++}$ in a single BS scenario. Therefore, we can use a gradient approach to compute the powers of the UEs to satisfy the SINDR constraint. Finally, the power of UE $k$ is updated iteratively in the negative direction of the gradient as
\begin{equation}
\begin{aligned}
\label{eq:gdUp}
\rho_k^{(i)} = \rho_k^{(i-1)} - \zeta \left. \frac{\partial \mathcal{L}({\rho_k, \mu_k})}{\partial \rho_k}\right \vert_{\{ \rho_{\bar k} = \rho_{\bar k}^{(i-1)}, \mu_{\bar k} =\mu_{\bar k}^{(i-1)} \}_{\bar k \in \setK}},
\end{aligned}
\end{equation}
where $\zeta \ge 0$ is the step size. The gradient of the Lagrangian with respect to $\rho_k$ is given in \eqref{eq:gradLag}--\eqref{eq:gradKbar} at the top of the next page. To compute this gradient, we need to evaluate the gradient of the matrix $\A$ with respect to $\rho_k$, which is given by
\addtocounter{equation}{+2}
\begin{align}
\frac{\partial}{\partial \rho_k}\A = -\sqrt{\frac{1}{2\pi}}\Diag(\C_{\y})^{-\frac{3}{2}}\Diag(\h_k\h_k^{\herm}).
\end{align}
Moreover, the gradient of the $(i,j)$th element of the real part of $\C_{\r}$ is given in \eqref{eq:gradCrReal} at the top of the next page, where we denote $q_{i,j} \triangleq \C_{\r}(i,j)$. The gradient for the imaginary part of $\C_{\r}$ can be obtained by replacing $\Re$ with $\Im$ in \eqref{eq:gradCrReal}. Finally, we have $\frac{\partial}{\partial \rho_k}\C_{r}(i,i) = 0,~\forall i$. The dual variable $\mu_k$ is updated as
\addtocounter{equation}{+1}
\begin{align}\label{eq:mu_up}
\mu_k^{(i)} = \mathrm{max}\bigg(0, \mu_k^{(i-1)} - \eta \bigg(\! \rho_k \h_k^{\herm}\A \C_{\r}^{-1} \A \h_k \! - \! \frac{\gamma_k^{\tar}}{\gamma_k^{\tar}+1}\bigg)\bigg),
\end{align}
where $\eta \ge 0$ is the step size and $i$ is the iteration index.

Note that the gradient update in \eqref{eq:gdUp} is terminated when the SINDR constraint is met for all UEs. However, the convergence speed of this method depends on the slope of the unimodal function at different UE transmit powers. Moreover, in the case of multiple BSs, this method may converge to a local optimum due to the presence of multiple peaks in the SINDR with respect to UE transmit power, as discussed in Section~\ref{sec:SINDRChar}.  

To meet the SINDR constraints of \eqref{eq:p2}, an appropriate UE transmit power can be determined using the standard fixed-point power update method~\cite{Wie06}. This method is effective at lower powers, where the SINDR resembles the SINR of an unquantized system.

\smallskip
\begin{figure}[t!]
\begin{minipage}[t]{0.935\linewidth}
\begin{algorithm}[H]
\footnotesize
\textbf{Data:} $\h_k$, $\{\gamma_k^{\tar}\}_{\{k \in \setK\}}$, $\epsilon > 0$. \\
\textbf{Initialization:} UE transmit power $\rho_k^{(0)}, \forall k$, iteration~$i=0$.
\begin{itemize}
\item Compute $\sinr_k^{(0)}$ as in \eqref{eq:sinr_mse_k}.
\end{itemize}
\While{$|\underset{k \in \setK}{\min}(\sinr_k^{(i)}) - \gamma_k^{\tar}| \ge \epsilon$}
{
    \begin{itemize}
    \item $i = i+1$.
    \item Update all the UE transmit powers as in \eqref{eq:pwrScal}.
    \item Compute the $\sinr_k^{(i)}$ as in \eqref{eq:sinr_mse_k}.
    \end{itemize}
       
        \If{ $\underset{k \in \setK}{\min}(\sinr_k^{(i)} \! - \! \sinr_k^{(i-1)} ) \! \le \! 0$}  
        {
          \textbf{break} \qquad \% $\gamma_k^{\tar}$ is not achievable.
        } 
  }
\caption{(Fixed-point)} \label{alg:pwrScal}
\end{algorithm}
\end{minipage}
\vspace{-2mm}
\end{figure}

\textit{\textbf{2) Fixed-Point Update.}} The transmit powers of the UEs are updated in parallel at each iteration. Specifically, at iteration~$i$, the transmit power of UE~$k$ is scaled to satisfy the SINDR constraint given by \eqref{eq:p2} using the fixed-point iterative power update method~\cite{Wie06}:
\begin{align}
\begin{aligned}
\label{eq:pwrScal} \rho_k^{(i)} = \frac{\gamma_k^{\tar}}{\sinr_k^{(i-1)}} \rho_k^{(i-1)},
\end{aligned}
\end{align}
where $\sinr_k^{(i-1)}$ is the SINDR of UE $k$ computed using the transmit powers $\{\rho_k^{(i-1)}\}_{k \in \setK}$ according to \eqref{eq:sinr_mse_k}. It is important to note that the power values must be initialized to the smallest possible value and cannot be arbitrarily large to ensure the feasibility of this method. As the SNDR is unimodal, unlike the unquantized system, we implement a termination criterion to terminate the power update process if the SINDR starts decreasing with the increase in transmit power. The fixed-point iterative power update method for computing the UE transmit powers is summarized in Algorithm \ref{alg:pwrScal}.

The fixed-point power update algorithm is more effective at lower powers where the SINDR increases linearly with the power values. However, at the optimal SINDR value, the increase in SINDR with respect to the UE transmit power is not linear, which may result in slower convergence. Additionally, since the interference and QD of 1-bit ADC do not follow the properties of a standard interference function, this algorithm may not have a unique solution in such cases.

\smallskip

\textit{\textbf{3) BCD Update.}} In the single BS case, the SINDR of a UE is unimodal with respect to the UE transmit power while keeping all other UE transmit powers fixed. Therefore, we employ a BCD method to update the UE transmit powers. In this method, we iteratively search for the required power of each UE to meet its target SINDR while keeping the other UE transmit powers fixed. The powers of all UEs are updated cyclically until the target SINDR is achieved by all UEs simultaneously, as described in Algorithm~\ref{alg:BCDpwrSea}. We use a simple line search with a  UE transmit power resolution of $\Delta \rho$ and maximum power of $\rho_{\ue}, \forall k$, to optimize the power values. However, a faster update of the power values can be achieved by using a backtracking line search since the SINDR function is unimodal. Moreover, the optimal power values in each iteration can be used as the initial values for the next iteration to expedite the updates.

Similar to the fixed-point algorithm, the BCD algorithm also terminates if the target SINDR is not achievable within the power limits or if the UE SINDR decreases with increasing the transmit power. This method can be extended to multi-BS scenarios, where the feasibility of noise levels at the BS needs to be ensured to achieve a better target SINDR. Since the SINDR of a UE is non-unimodal for any noise level at the BS, we evaluate the algorithm for different noise levels at each BS to find the first peak in the SINDR. Then, we select the optimal SINDR value among all the computed noise levels.

\subsection{Power Control Based on Max-Min-SINDR}

In this section, we aim to optimize the UE transmit powers subject to a power constraint while maximizing the minimum SINDR objective, which can be formulated as
\begin{align} \label{eq:p3}
\begin{array}{cl}
\underset{\{\rho_{k}, \sigma_b\}}{\maximize} & \underset{k \in \setK}{\min} \ \sinr_k \\
\mathrm{s.t.} &  \rho_k \le \rho_{\ue}, \quad \forall k \\
 & \sigma_{b} \ge \sigma_{\min}. \quad \forall b
\end{array}
\end{align}
Continuing with a similar approach as outlined in Section~\ref{subsec:pcmp}, the optimization of UE transmit powers, given fixed $\{\sigma_{b}\}_{b \in \setB}$ values, can be achieved through the following simplified epigraph optimization problem:\footref{note1}%
\begin{align} \label{eq:p4}
\begin{array}{cl}
\underset{\{\rho_{k}\}, \gamma}
{\maximize} & \quad \gamma \\
\mathrm{s.t.} & \quad \rho_k \h_k^{\herm}\A \C_{\r}^{-1} \A \h_k \ge \frac{\gamma}{\gamma+1} , \quad \forall k \\
& \quad \rho_k \le \rho_{\ue}, \quad \forall k
\end{array}
\end{align}
where $\gamma \triangleq \underset{k}{\min} \quad  \! \! \! \sinr_k$. The above problem can be solved using a bisection algorithm over $\gamma$. For a fixed value of $\gamma$ (the same for all the UEs), we can solve \eqref{eq:p2} to check the feasibility of \eqref{eq:p4}. Therefore, we can use the same algorithms, such as the gradient, fixed-point, and BCD, to optimize the UE transmit powers for the max-min SINDR problem. Note that at the optimal $\gamma$ value, it is possible that none of the power constraints meet equality. However, in the unquantized system, at least one of the power constraints meets equality at the optimal value of $\gamma$.

\begin{figure}[t!]
\begin{minipage}[t]{0.815\linewidth}
\begin{algorithm}[H]
\footnotesize
\textbf{Data:} $\h_k$, $\rho_{\ue}$, $\gamma_k^{\tar}$, $\epsilon > 0$, $\Delta \rho$. \\  
\textbf{Initialization:} UE transmit power $\rho_k^{(0)}, \forall k$, iteration~$j=0$, $p = 0$. \\
\While{$|\underset{k \in \setK}{\min}(\sinr_k^{(j)}) - \gamma_k^{\tar} | \ge \epsilon$}
{
    \begin{itemize}
        \item $j = j+1$.
    \end{itemize}
    \For{$k=1, \ldots, K$}
    {
        \begin{itemize}
            \item $p = 0$, $\sinr_k^{(0)} = 0$. 
        \end{itemize}
        \For{$\rho = \rho_k^{(j-1)} : \Delta \rho : \rho^{\ue}$}
        {
            \begin{itemize}
                \item $p = p +1 $.
                \item Compute the $\sinr_k^{(p)}$ of UE~$k$  as in \eqref{eq:sinr_mse_k} \\ with the UE transmit powers of $\qquad \qquad \qquad \qquad $ $[\rho_1^{(j)}, \ldots, \rho_{k-1}^{(j)}, \rho, \rho_{k+1}^{(j-1)},\ldots,\rho_K^{(j-1)}]$.
            \end{itemize}
          \If{$|\sinr_k^{(p)} - \gamma_k^{\tar}|\le \epsilon$ \textnormal{or} $\sinr_k^{(p)} \le \sinr_k^{(p-1)} $}
          {
             \textbf{break}
          }
        }
        \begin{itemize}
            \item $\rho_k^{(j)} = \rho$.
        \end{itemize}
        \If{$|\sinr_k^{(p)} - \gamma_k^{\tar}| > \epsilon$}
          {
             \textbf{break}
          }
    }
    \If{$|\sinr_k^{(p)} - \gamma_k^{\tar}|> \epsilon$ } 
        {
          \textbf{break} \qquad \% $\gamma_k^{\tar}$ is not achievable.
        }
    \begin{itemize}
        \item Compute the $\sinr_k^{(j)}$ of all the UEs with the updated power values, i.e., $\{\rho_k^{(j)}\}_{k \in \setK}$.
    \end{itemize}
  }
\caption{(BCD)} \label{alg:BCDpwrSea}
\end{algorithm}
\end{minipage}
\vspace{-3mm}
\end{figure}

\vspace{-1mm}
\section{Numerical Results} \label{sec:NUM}
\vspace{-1mm}
\begin{figure*}
\vspace{1mm}
    \begin{minipage}[t]{0.31\textwidth}
        \begin{figure}[H]
            \begin{tikzpicture}

\begin{axis}[
	width=6.15cm,
	height=5.15cm,
	xmin=0, xmax=15.5,
	ymin=5, ymax=25,
    xlabel={Target $\mathrm{SINDR}$},
    ylabel={Min-power [dBm]},
    xlabel near ticks,
	ylabel near ticks,
    x label style={font=\footnotesize},
	y label style={font=\footnotesize},
    ticklabel style={font=\footnotesize},
    legend pos=south east,
    legend cell align=left,
    legend columns=1,
    legend style={font=\scriptsize, inner sep=1pt, fill opacity=0.8, draw opacity=1, text opacity=1},
	grid=both,
]

\addplot[line width=1pt, black]
table[x=x_gd_1bs, y=y_gd_1bs, col sep=comma] 
{Figures/Data/Fig4.txt};
\addlegendentry{Gradient};

\addplot[line width=1pt, red, dashed, mark=o, mark options={solid}]
table[x=x_fp_1bs, y=y_fp_1bs, col sep=comma] 
{Figures/Data/Fig4.txt};
\addlegendentry{Fixed-point};

\addplot[line width=1pt, blue, dotted, mark=star, mark options={solid}]
table[x=x_bcd_1bs, y=y_bcd_1bs, col sep=comma] 
{Figures/Data/Fig4.txt};
\addlegendentry{BCD};




\end{axis}

\end{tikzpicture}
            \vspace{-1mm}
            \caption{\small{Single-BS case: min-power vs target SINDR.}} \label{fig:UEtxvsSINDR}
        \end{figure}
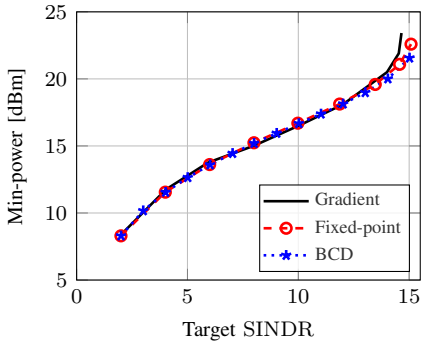
    \end{minipage}
    \hfill
    \begin{minipage}[t]{0.31\textwidth}
        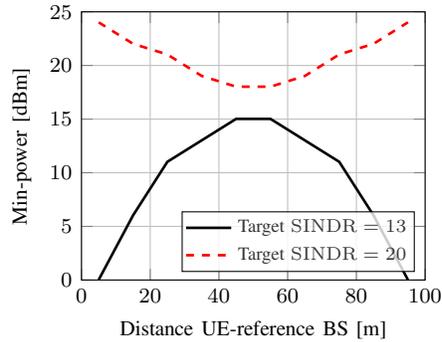
\begin{figure}[H]
            \begin{tikzpicture}

\begin{axis}[
	width=6.15cm,
	height=5.15cm,
	xmin=0, xmax=100,
	ymin=0, ymax=25,
    xlabel={Distance UE-reference BS [m]},
    ylabel={Min-power [dBm]},
	ytick={0,5,10,15,20,25},
    xlabel near ticks,
	ylabel near ticks,
    x label style={font=\footnotesize},
	y label style={font=\footnotesize},
    ticklabel style={font=\footnotesize},
    legend pos=south east,
    legend cell align=left,
    legend columns=1,
    legend style={font=\scriptsize, inner sep=1pt, fill opacity=0.8, draw opacity=1, text opacity=1},
	grid=both,
]

\addplot[line width=1pt, black]
table[x=dist, y=sindr13, col sep=comma] 
{Figures/Data/Fig5.txt};
\addlegendentry{Target $\mathrm{SINDR} = 13$};

\addplot[line width=1pt, red, dashed]
table[x=dist, y=sindr20, col sep=comma] 
{Figures/Data/Fig5.txt};
\addlegendentry{Target $\mathrm{SINDR} = 20$};

\end{axis}

\end{tikzpicture}
             \vspace{-1mm}
            \caption{\small{Two-BS case: min-power vs distance UE-reference BS.}} \label{fig:DisvsPwr}
        \end{figure}
    \end{minipage}
    \hfill
    \begin{minipage}[t]{0.31\textwidth}
        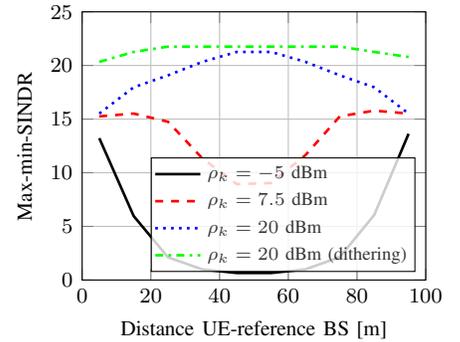
\begin{figure}[H]
            \begin{tikzpicture}

\begin{axis}[
	width=6.15cm,
	height=5.15cm,
	xmin=0, xmax=100,
	ymin=0, ymax=25,
    xlabel={Distance UE-reference BS [m]},
    ylabel={Max-min-SINDR},
	ytick={0,5,10,15,20,25},
    xlabel near ticks,
	ylabel near ticks,
    x label style={font=\footnotesize},
	y label style={font=\footnotesize},
    ticklabel style={font=\footnotesize},
    legend pos=south east,
    legend cell align=left,
    legend columns=1,
    legend style={font=\scriptsize, inner sep=1pt, fill opacity=0.8, draw opacity=1, text opacity=1},
	grid=both,
]

\addplot[line width=1pt, black]
table[x=Dist, y=y_p_5dbm, col sep=comma] 
{Figures/Data/Fig6.txt};
\addlegendentry{$\rho_k = -5$~dBm};

\addplot[line width=1pt, red, dashed]
table[x=Dist, y=y_p7p5dBm, col sep=comma] 
{Figures/Data/Fig6.txt};
\addlegendentry{ $\rho_k = 7.5$~dBm};

\addplot[line width=1pt, blue, dotted]
table[x=Dist, y=y_20dBm_notun, col sep=comma] 
{Figures/Data/Fig6.txt};
\addlegendentry{$\rho_k = 20$~dBm};

\addplot[line width=1pt, green, dash dot]
table[x=Dist, y=y_p20dBm, col sep=comma] 
{Figures/Data/Fig6.txt};
\addlegendentry{$\rho_k = 20$~dBm (dithering)};

\end{axis}

\end{tikzpicture}
             \vspace{-1mm}
            \caption{\small{Two-BS case: max-min-SINDR vs distance UE-reference BS.}} \label{fig:DisvsSINDR}
        \end{figure}
    \end{minipage}
    \vspace{-3mm}
\end{figure*}

To evaluate the performance of the system we consider the similar setup as  discussed in Section~\ref{sec:SINDRChar}, with four single-antenna UEs placed equidistant from both the BSs. The AWGN power at the BSs is set to $\sigma^2_{\min}=-95$~dBm, and dithering (noise level) can be added on top of it if needed to improve the performance. Fig.\ref{fig:UEtxvsSINDR} illustrates the min-power as a function of the target SINDR in a single-BS case. The \textit{Gradient}, \textit{Fixed-point}, and \textit{BCD} update methods perform similarly. However, due to the QD of the 1-bit ADC BS, the maximum feasible target SINDR is limited to around $15$. The computational complexity of the \textit{Gradient} method in each iteration is determined by \eqref{eq:gradLag}, which is higher compared to the \textit{Fixed-point}, and \textit{BCD} methods. Convergence is also influenced by the step sizes, represented as $\zeta$ and $\eta$ in \eqref{eq:gdUp} and \eqref{eq:mu_up}, respectively. On the other hand, the \textit{Fixed-point} method has lower computational complexity than the \textit{BCD} method. However, when approaching the optimal SINDR, the \textit{BCD} method may exhibit superior performance owing to the fixed increment of power in each step, i.e., $\Delta\rho$. Consequently, in the subsequent scenarios involving two-BSs, we employ the \textit{BCD} method to characterize the min-power and max-min SINDR designs.


Fig.~\ref{fig:DisvsPwr} shows the min-power required to achieve a given target SINDR as a function of the distance between the UE and the reference BS. In this simulation, two BSs are located $100$~m apart, and the UEs move along the line connecting both BSs. If the target SINDR is achieved by a single BS, such as SINDR of $13$, the required power increases as the UE moves away from the BS, as expected in the unquantized system. Therefore, the maximum min-power is required when the UE is located between the two BSs. However, if the target SINDR requires joint reception from both BSs with dithering, as in the case of SINDR of $20$, the min-power required to achieve the target SINDR is maximum when the UE is closer to one of the BSs and minimum at the cell edge, which is $50$~m in this case. This result differs from the unquantized system.

Considering the same setup used for Fig.~\ref{fig:DisvsPwr}, in Fig.~\ref{fig:DisvsSINDR}, we  characterized the max-min-SINDR as a function of the UE distance from the reference BS for various UE transmit powers. When the UE transmit power is too low, such as $-5$~dBm, the achievable SINDR decreases as the UEs move further from the BS, and it is lowest at the cell edge ($50$~m), as expected in the unquantized system. If we increase the UE transmit power constraint to $7.5$~dBm, the SINDR is lower when the UEs are closer to a BS since there is no combining gain from the other BS. The SINDR is maximum at $15$~m, where both the UE transmit power and combining gains are optimal, and it decreases as the distance from the BSs increases due to the limited UE transmit power. In the aforementioned scenarios with limited UE transmit power, the presence of dithering at the BSs may not improve the SINDR. However, when the UE transmit power is higher, such as $20$~dBm, the UE is not constrained by transmit power even at the cell edge. This allows for the maximum achievable SINDR by combining across the BSs, irrespective of the presence of dithering. As UEs move away from the cell edge, the SINDR decreases without dithering. Nevertheless, incorporating dithering significantly improves and maintains the maximum SINDR, as long as the UEs have sufficient power to reach the farthest serving BS, as depicted in Fig.~\ref{fig:DisvsSINDR}. However, UEs located near a BS still lack the necessary power to reach other BS, resulting in a decrease in the achievable SINDR. 

\vspace{-1mm}
\section{Conclusions}
\vspace{-1mm}
We investigated the uplink SNDR of a UE in both single-BS and multi-BS scenarios with 1-bit ADCs. In a single-BS case, the SNDR is unimodal, while it is non-unimodal in a multi-BS case. However, by adjusting the dithering at the BSs, the SNDR can be optimized and made unimodal. We optimized the UE transmit powers using the min-power and max-min-SINDR criteria and employed gradient, fixed-point, and BCD methods to solve the optimal UE transmit powers. The numerical analysis shows that in a single-BS system, the SINDR saturates beyond a specific UE transmit power. Meanwhile, in a multi-BS system, when the target SINDR is attainable through reception from multiple BSs with dithering, the UE transmit power depends on the farthest serving BS. In this context, we have demonstrated that incorporating dithering at the nearest BSs can effectively enhance the SINDR for UEs with a large disparity in path loss among the serving BSs.

\bibliographystyle{IEEEtran}
\bibliography{refs_abbr,refs}
\end{document}